\documentclass[twocolumn,amssymb,amsmath,pra, floatfix, superscriptaddress]{revtex4-1}

\usepackage{physics}
\usepackage{graphicx} 
\usepackage[caption=false, labelformat=simple]{subfig}
\usepackage{dcolumn}
\usepackage{csquotes}
\usepackage[colorlinks=true,linkcolor = {blue}, citecolor = {blue}, urlcolor = {blue}]{hyperref}
\usepackage{xcolor}

\begin{document}

\title{Efimovian three-body potential from broad to narrow Feshbach resonances}

\author{J. van de Kraats}
\email{j.v.d.kraats@tue.nl}
\affiliation{Eindhoven University of Technology, P. O. Box 513, 5600 MB Eindhoven, The Netherlands}
\author{D.J.M. Ahmed-Braun}
\affiliation{Eindhoven University of Technology, P. O. Box 513, 5600 MB Eindhoven, The Netherlands}
\author{J, -L. Li}
\affiliation{Eindhoven University of Technology, P. O. Box 513, 5600 MB Eindhoven, The Netherlands}
\affiliation{Institut f{\"u}r Quantenmaterie and Center for Integrated Quantum Science and Technology IQ ST, Universit{\"a}t Ulm, D-89069 Ulm, Germany}
\author{S.J.J.M.F. Kokkelmans}
\affiliation{Eindhoven University of Technology, P. O. Box 513, 5600 MB Eindhoven, The Netherlands}
\date{\today}

\begin{abstract}
    We analyse the change in the hyperradial Efimovian three-body potential as the two-body interaction is tuned from the broad to narrow Feshbach resonance regime. Here, it is known from both theory and experiment that the three-body dissociation scattering length $a_-$ shifts away from the universal value of $-9.7 \ r_{\mathrm{vdW}}$, with $r_{\mathrm{vdW}} = \frac{1}{2} \left(m C_6/\hbar^2 \right)^{1/4}$ the two-body van der Waals range. We model the three-body system using a separable two-body interaction that takes into account the full zero-energy behavior of the multichannel wave function. We find that the short-range repulsive barrier in the three-body potential characteristic for single-channel models remains universal for narrow resonances, whilst the change in the three-body parameter originates from a strong decrease in the potential depth. From an analysis of the underlying spin structure we further attribute this behavior to the dominance of the two-body interaction in the resonant channel compared to other non-resonant interactions.
\end{abstract}
\maketitle

\section{INTRODUCTION}
\label{sec:intro}

In his seminal papers \cite{Efimov1970, Efimov1971}, Vitaly Efimov predicted the appearance of an infinite and geometrically spaced set of three-particle bound states as the pairwise interaction becomes resonant. These Efimov states are bound by a universal attractive potential, decaying asymptotically as $-1/R^2$ for three particles at root mean square separation $R$. In trapped ultracold atomic gases the Efimov effect induces log-periodic peaks in the atom loss rate, driven by enhanced three-body recombination when an Efimov trimer crosses into the three-particle continuum \cite{Esry1999, Kraemer2006, Braaten2006, Naidon2017}. The position of the loss peak associated with the ground Efimov state sets a characteristic length scale $a_-$, commonly referred to as the three-body parameter. In three-body systems with zero-range interactions, introducing a three-body parameter is neccessary to regularise the scale invariant unbounded Efimov spectrum \cite{Braaten2006}.

Despite its short-range nature, experiment has revealed that the three-body parameter in different atomic species attains a value close to $a_- = -9.7 \ r_{\mathrm{vdW}}$ \cite{ Berninger2011, Gross2011, Dyke2013, Wild2012, Chapurin2019}, where $r_{\mathrm{vdW}} = \frac{1}{2} \left(m C_6/\hbar^2 \right)^{1/4}$ is the van der Waals length associated with the long-range two-body interaction. Subsequent theoretical studies have found that this \textquote{van der Waals universality} originates from a characteristic suppression of the two-body wave function when $r< r_{\mathrm{vdW}}$, where $r$ is the two-particle separation \cite{Wang2012, Naidon2014}. This suppression leads to the appearance of a strong repulsive barrier in the three-body potential at mean square separations $R \approx 2 \ r_{\mathrm{vdW}}$, which shields the particles from probing the non-universal short-range detail of the atomic species.

The above-mentioned theoretical analyses are based on single-channel interaction potentials, which are expected to be accurate provided that the intrinsic length scale $r_*$ due to the resonance width is much smaller than the potential range. This broad resonance regime may be defined by a large resonance strength parameter $s_{\mathrm{res}} = \bar{a}/r_* \gg 1$ \cite{Chin2010}, where $\bar{a} \approx 0.955978 \ r_{\mathrm{vdW}}$ is the mean scattering length of the van der Waals interaction \cite{Gribakin1992}. The opposite case of a narrow resonance, where $s_{\mathrm{res}} \ll 1$, is characterized by universal behavior in terms of the dominant length scale $r_* \gg r_{\mathrm{vdW}}$. In this limit, treatments of the three-body problem which neglect the details of the van der Waals interaction have found the three-body parameter to be determined universally as $a_- = -10.9 \ r_*$ \cite{Petrov2004, Nishida2012, Naidon2017, Secker2021_2}. Connecting the broad and narrow resonance limits through the intermediate regime where $s_{\mathrm{res}} \approx 1$ with a van der Waals interaction model remains to be desired, in particular given that recent experiments in this regime in \textsuperscript{39}K have revealed clear deviations from both universal limits \cite{Chapurin2019}. A key aspect of this problem is the change in structure of the trimer and its associated potential energy surface as a function of the resonance strength, which will be the central topic of this paper.

In this work we study the Efimovian three-body potential using a realistic multichannel two-body van der Waals interaction, which can be easily tuned to probe a wide regime of resonance strengths. To solve the three-body problem we approximate this interaction by a separable potential which reproduces the zero-energy wave function of the original interaction. We then derive an effective three-body potential from the open-channel three-body wave function, which models the actual three-body potential that binds the Efimov state. Subsequently we study the dependence of this potential on the resonance strength $s_{\mathrm{res}}$, and provide an analysis of our findings in terms of the multichannel structure underlying the three-body dynamics. 

This paper will be structured as follows. In Sec. \ref{sec:2body} we outline our approach at the two-body level, first defining a two-channel model interaction with a Feshbach resonance that can be tuned from the broad to narrow resonance strength limit. Subsequently we formulate a separable approximation to this interaction. In Sec. \ref{sec:3body} we move to the three-body level, which we analyse first in momentum space to facilitate our actual computations, and then subsequently in position space for our analysis of the three-body potential. In Sec. \ref{sec:res} we present and analyse our results, after which we conclude this paper in Sec. \ref{sec:conc}.

\section{Two-body interaction models}
\label{sec:2body}

\subsection{Model two-channel interaction}
\label{sec:2chan}

In this section we develop a flexible two-channel model that can be tuned to produce a Feshbach resonance with a given Breit-Wigner shape \cite{Mies2000, Nygaard2006, Wang2014}. We label the two scattering channels $\sigma = \left\{1,2 \right\}$, with internal energies $\varepsilon_{\sigma}$, and define the two-body Hamiltonian operator,
\begin{align}
\begin{split}
H &= H^0 + V = \begin{pmatrix}
H_1^0 + V_{1,1} & V_{1,2} \\ 
V_{2,1} & H_2^0 + V_{2,2}
\end{pmatrix}. 
\end{split}
\end{align}
where $H^0$ contains the internal and kinetic energies of the particles, and $V$ the pairwise interactions. Expressed in the interparticle distance $r$, the diagonal interactions are of van der Waals type \cite{LJPexp, Pade2007},
\begin{align}
\begin{split}
V_{1,1}(r) = V_{2,2}(r) =  C_6 \left(\frac{r_0^4}{r^{10}} - \frac{1}{r^6} \right),
\end{split}
\label{eq:Vo}
\end{align}
with $C_6$ the species specific dispersion coefficient. The parameter $r_0$ controlling the short-range barrier is tuned such that both channels have 8 uncoupled dimer states, which we have confirmed is sufficiently deep such that the scattering is universally determined by the van der Waals tail. The off-diagonal terms of the interaction represent spin-exchange processes, which we model using a Gaussian form inspired by Ref.~\cite{Wang2014},
\begin{align}
\begin{split}
V_{1,2}(r) = V_{2,1}(r) = \beta e^{-\alpha(r-r_{W})^2},
\end{split}
\label{eq:2chancouple}
\end{align}
where $\left\{\beta,\alpha, r_W\right\}$ are tuneable parameters. To enforce the short-range nature of the spin-exchange interaction, we fix $r_W = 0.15 \ r_{\mathrm{vdW}}$ \cite{rw}. We will take the channel $\sigma = 1$ to be energetically open, and set its internal energy as $\varepsilon_1 = 0$ such that $H_1^0(r) = -\hbar^2 \nabla_r^2/m$. The channel $\sigma=2$, referred to as the closed channel, has a magnetic field dependent internal energy, $H_2^0(r, B) = -\hbar^2 \nabla_r^2/m + \varepsilon_2(B)$. To model the Feshbach resonance, we define \cite{Mies2000},
\begin{align}
\begin{split}
\varepsilon_{2}(B) = \varepsilon_{\mathrm{b}} + \delta \mu \left(B - B_{\mathrm{res}} \right),
\end{split}
\end{align}
where $\varepsilon_{\mathrm{b}}$ is the bare binding energy of the resonant bound state in the closed channel, $\delta \mu$ the differential magnetic moment of the particles which is inferred from experiment, and $B_{\mathrm{res}}$ the bare resonant magnetic field. Given a background scattering length $a_{\mathrm{bg}}$, resonance width $\Delta B$ and resonant magnetic field $B_0$, we can extract an associated set of model parameters $\left\{r_0, \alpha, \beta, r_W, B_{\mathrm{res}} \right\}$. The details of this mapping are outlined in Appendix \ref{ap:2bodytuning}. The resulting resonance strength is obtained as \cite{Chin2010},
\begin{align}
\begin{split}
s_{\mathrm{res}} = \frac{m}{\hbar^2} \bar{a} a_{\mathrm{bg}} \delta \mu \Delta B,
\end{split}
\end{align}
which quantifies the ratio $r_*/\bar{a}$ as mentioned in Sec. \ref{sec:intro}.
\subsection{EST separable potential}
\label{sec:EST}

As pointed out in previous studies, the universal van der Waals three-body parameter and three-body potential can be reproduced by accounting for the full finite-range detail of the van der Waals interaction \cite{Wang2012}. Similarly it was recently shown that reproducing the three-body recombination rate for resonances of intermediate strength in \textsuperscript{39}K requires an inclusion of the exact three-body spin structure in the Hamiltonian \cite{Secker2021}. Such approaches however, are complicated numerically, and not conducive to our goal of developing a simple and flexible model. Fortunately it was pointed out in Refs. \cite{Naidon2014, Naidon2014_2} that van der Waals universality can be reproduced using a much simpler model, based on the Ernst, Shakin and Thaler (EST) separable potential \cite{Ernst1973}. In this section we develop such an approach for our multichannel interaction. The crucial point is that we approximate the interaction in such a way that the full two-body wave function at zero energy is taken into account, whilst retaining the simplicity of a single-term separable potential.

\begin{figure}[t]
\includegraphics[height=0.22\textwidth]{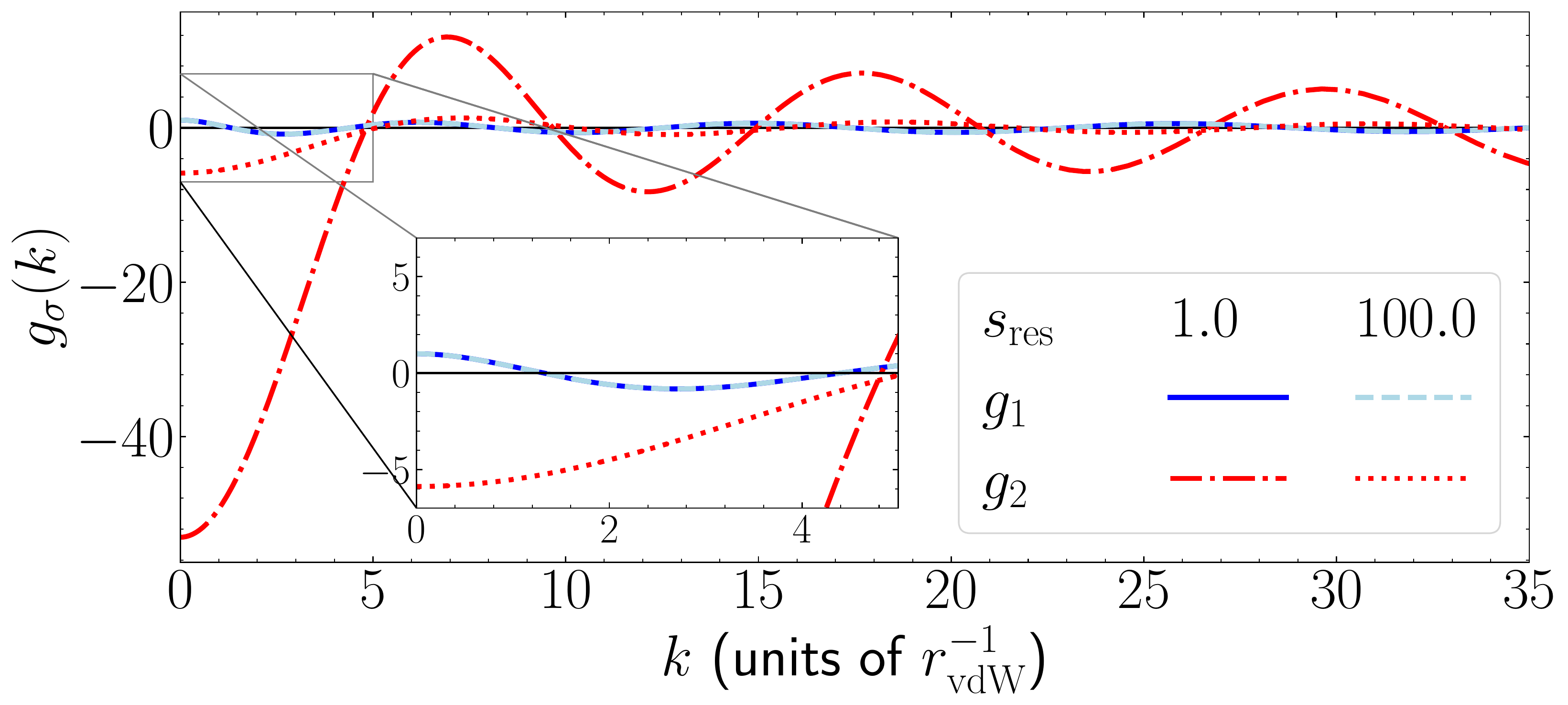}
\caption{\label{fig:gk}The form factors $g_{\sigma}(k)$ as a function of momentum $k$, tuned to two different resonance strengths. Inset shows a zoom of the low momentum regime, where one observes the normalisation $g_{1}(0) = 1$.}
\end{figure}

For an arbitrary multichannel interaction $V$, one may define a separable approximation as $V^{\mathrm{sep}} = \ket{g} \xi \bra{g}$.
In the EST formalism, the form factor $\ket{g}$ and potential strength $\xi$ are derived from a given eigenfunction $\ket*{\psi}$ of the full multichannel Hamiltonian,
\begin{align}
\ket*{g} = V \ket*{\psi}, \qquad \xi^{-1} = \matrixel*{\psi}{V}{\psi}.
\label{eq:ESTdef}
\end{align}
With these definitions one may show that $\ket{\psi}$ is also an eigenfunction of the Hamiltonian where $V$ is replaced with $V^{\mathrm{sep}}$, with the exact same eigenvalue \cite{Ernst1973}. Adopting the approach of Ref.~\cite{Naidon2014}, we take $\ket{\psi}$ to be the zero-energy scattering state, such that our model takes as input the low-energy scattering detail of the actual interaction. The separable interaction has an associated separable $t$-matrix, given by the Lipmann-Schwinger equation \cite{Taylor2006},
 \begin{align}
t^{\mathrm{sep}}(z) = V^{\mathrm{sep}} + V^{\mathrm{sep}} G_0(z)t^{\mathrm{sep}}(z).
\label{eq:LSt}
\end{align}
Here $G_0(z) =(z - H_0)^{-1}$ is the Green's function in the absence of interactions. We define its s-wave eigenstates as $\ket*{k,\sigma}$, where $k = \abs*{\vb{k}}$ is the relative momentum and $\sigma$ the scattering channel introduced in the previous section. In this basis, the transition matrix may be written as,
\begin{align}
t^{\mathrm{sep}}_{\sigma',\sigma}(z, k', k) = g_{\sigma'}(k') \tau(z) g_{\sigma}^*(k),
\label{eq:tsep}
\end{align}
where $t^{\mathrm{sep}}_{\sigma',\sigma}(z, k', k) = \matrixel*{k',\sigma'}{t^{\mathrm{sep}}(z)}{k,\sigma}$ and $g_{\sigma}(k) = \braket*{k,\sigma}{g}$. Explicit expressions for $\tau(z)$ and $g_{\sigma}(k)$ are given in Appendix \ref{ap:EST}. To obtain the eigenfunction $\ket*{\psi}$ we explicitly diagonalise the two-body Hamiltonian, using a mapped grid discrete variable representation \cite{Willner2004, Secker2021_3}. For the broad resonance limit $s_{\mathrm{res}} \gg 1$, the Feshbach resonance is well approximated by a potential resonance in a single-channel model with the interaction in Eq. \eqref{eq:Vo}. In this case, we obtain $\ket*{\psi}$ using an efficient Numerov method \cite{Karman2014}. The behavior of the form factor as a function of resonance width is illustrated in Fig.~\ref{fig:gk}. Note that the arbitrary normalisation of the form factors is fixed by taking $g_1(0) = 1$. Since the open-channel component of the wave function is independent of $s_{\mathrm{res}}$ (see Sec. \ref{sec:repuls}), the open-channel form factors are much less sensitive to changes in the resonance strength than the closed-channel form factors.

\begin{figure}[t]
\includegraphics[height=0.35\textwidth]{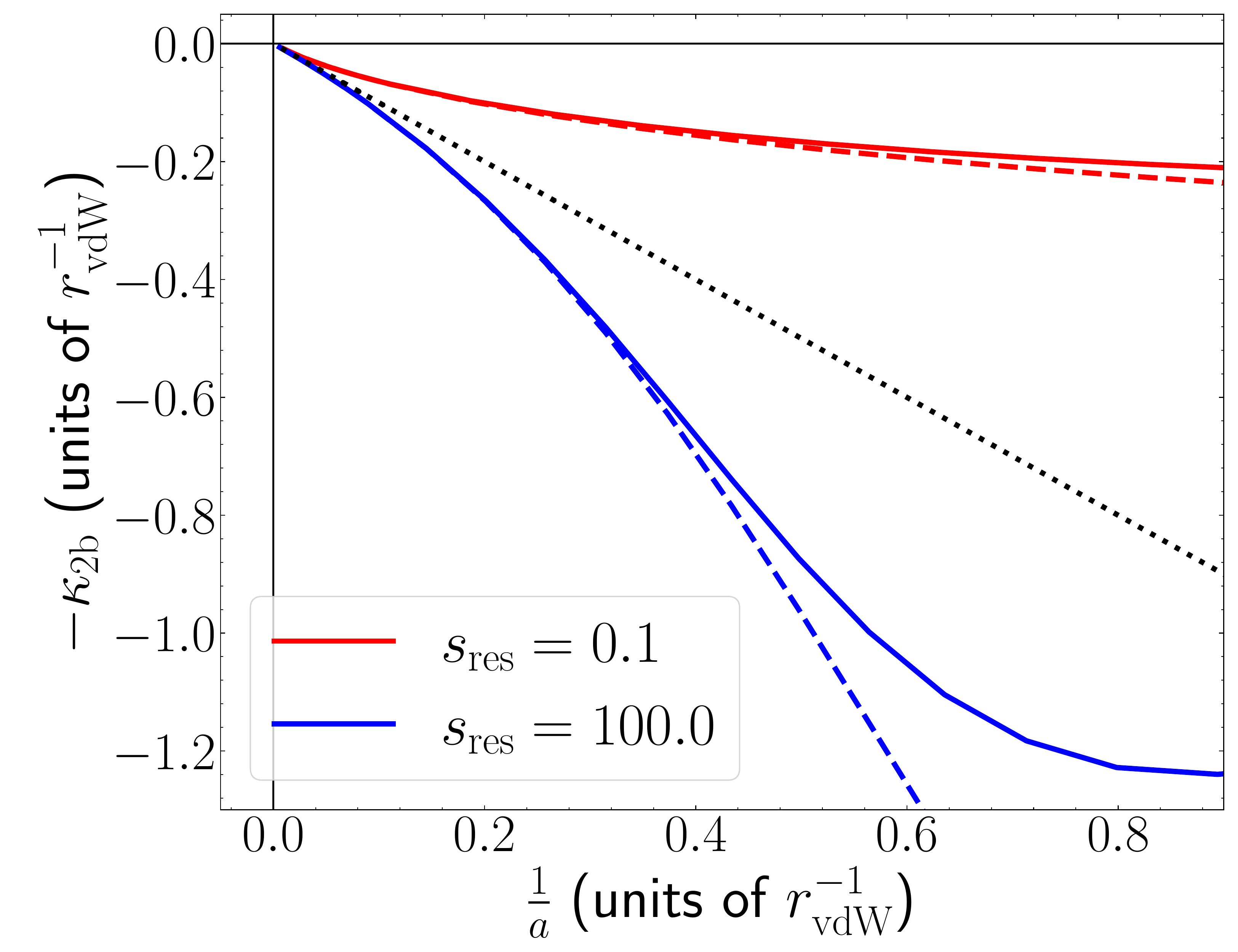}
\caption{\label{fig:EfPlot_dim} Binding wavenumber $\kappa_{\mathrm{2b}} = \sqrt{m \abs*{\varepsilon}/\hbar^2}$ of the Feshbach dimer that manifests in the two-channel model of Sec. \ref{sec:2chan}, as a function of inverse scattering length. Results are shown both with the EST model potential (solid lines), and a direct solution of the multichannel Schr\"odinger equation (dashed lines). We show results for two different resonance strengths, and additionally plot the universal resonant result $\kappa_{\mathrm{2b}} \sim 1/a$ as the black dotted line.}
\end{figure}

We emphasize that the EST model is based on the zero-energy wave function, and hence loses accuracy when used to describe deep bound states. To illustrate this behavior we have computed the shallow dimer energy around the 8th potential resonance in the two-channel model, both by a direct numerical solution of the multichannel Schr\"odinger equation, and via the EST potential of this section. In the latter case a dimer solution is found through the condition $\tau^{-1}(\varepsilon) = 0$, with $\varepsilon < 0$ the binding energy. The two results are compared in Fig.~\ref{fig:EfPlot_dim}, where one observes that the EST potential is most accurate near threshold, and is thus naturally suited to treat states near resonance. For smaller scattering lengths the EST potential becomes inaccurate for broad resonances where the dimer becomes too strongly bound, but remains reasonably accurate in describing narrower resonances. In this paper we only concern ourselves with the near resonant regime $a \gg r_{\mathrm{vdW}}$, where the EST potential is accurate regardless of the resonance strength. 

\section{Three-body approach}
\label{sec:3body}

\subsection{Three-body integral equation}
\label{sec:ESC_EST}

To find the three-body bound state energy $E$ and wave function $\ket*{\Psi}$, we adopt the methodology of Ref. \cite{Secker2021_2}. Three-body states are expressed in the $s$-wave basis $\ket*{k,p, \sigma, \sigma_3}$. Here $\left\{k, p \right\}$ are the dimer and atom-dimer Jacobi momenta respectively, $\sigma = \underline{\sigma_1 \sigma_2}$ gives the channel of the two-body system as introduced in Sec. \ref{sec:2body}, where the underline denotes symmetrization, and $\sigma_3$ the spin of the third particle (in general in this section we shall use subscripts to label particles $1,2$ and $3$). The internal energy of the three particles as dictated by the spin states will be denoted as $E(\sigma, \sigma_3)$. The wave function $\ket*{\Psi}$ is  written into the Faddeev decomposition \cite{Faddeev1960, Glockle1983},
\begin{align}
\begin{split}
\ket*{\Psi} = (1 + P_+ + P_-)\ket*{\bar{\Psi}},
\end{split}
\label{eq:FaddeevDecomp}
\end{align}
with $\ket*{\bar{\Psi}}$ the Faddeev component and $P_{\pm}$ cyclic permutation operators of the particle indices. For the EST separable two-body transition matrix (see Eq. \eqref{eq:tsep}), the state $\ket*{\bar{\Psi}}$ projected on our three-body basis can be formulated as,
\begin{align}
\begin{split}
\braket*{k, p, \sigma, \sigma_3}{\bar{\Psi}} = \frac{g_{\sigma}(k) F_{\sigma_3}(p)}{E - \frac{\hbar^2 k^2}{m} - \frac{3}{4} \frac{\hbar^2 p^2}{m} - E(\sigma, \sigma_3)}.
\end{split}
\label{eq:FadComp}
\end{align}
Here the function $F_{\sigma_3}(p)$ captures the dynamics of the third particle, and is obtained from the one-dimensional integral equation,
\begin{align}
\begin{split}
&\tau^{-1}\left( Z_{\sigma_3}(p) \right) F_{\sigma_3}(p) = \sum_{\sigma_3'} \int d^3\vb{q} \ \mathcal{Z}_{\sigma_3 \sigma_3'}(\vb{p},\vb{q})  F_{\sigma_3'}(q),
\end{split}
\label{eq:STM1}
\end{align}
known as the Skorniakov-Ter-Martirosian (STM) equation. Here $Z_{\sigma_3}(\kappa,p) = E - 3\hbar^2p^2/(4m) - E_{\sigma_3}$, with $E_{\sigma_3}$ the internal energy of the third particle, and the kernel $\mathcal{Z}_{\sigma_3 \sigma_3'}(\vb{p},\vb{q})$ is given as,
\begin{align}
\begin{split}
&\mathcal{Z}_{\sigma_3 \sigma_3'}(\vb{p},\vb{q}) =  \sum_{\sigma \sigma'}\frac{g_{\sigma}\left(\abs*{\vb{q} + \frac{1}{2} \vb{p}} \right) g_{\sigma'}^*\left(\abs*{\frac{1}{2}\vb{q} +  \vb{p}} \right)}{E - \frac{\hbar^2}{m}(p^2 + q^2 + \vu{p} \cdot \vu{q}) - E(\sigma', \sigma_3')} \\ & \qquad \qquad \qquad \times\matrixel*{\sigma, \sigma_3}{2P_+^s}{\sigma', \sigma_3'}.
\end{split}
\label{eq:STM2}
\end{align}
Here $P_+^s$ is the cyclic permutation operator that acts solely in the space of internal states, and the factor $2$ accounts for the anticyclic permutation. We will label distinct single particle states in alphabetical order, taking the open two-body channel as $\ket*{\sigma=1} \equiv \ket*{\underline{aa}}$, and the closed-channel as $\ket*{\sigma=2} \equiv \ket*{\underline{bc}}$. The presence of $P_+^s$ in the integral equation expresses that the three-body problem is complicated by the appearance of the additional non-resonant closed channels $\ket*{\underline{ab}, c}$ and $\ket*{\underline{ac},b}$. However, the non-resonant interaction in $\ket*{\underline{ab}}$ and $\ket*{\underline{ac}}$ is much weaker than the resonance enhanced interaction in $\ket*{\underline{bc}}$, which we illustrate numerically in Appendix \ref{ap:comp}. Hence, we choose to neglect $V_{\underline{ab}, \underline{ab}}$ and $V_{\underline{ac}, \underline{ac}}$, which restricts the multichannel three-body equation \eqref{eq:STM1} to the channels $\ket*{\underline{aa}, a}$ and $\ket*{\underline{bc},a}$. This is equivalent to the fixed spectating spin approximation, which is widely implemented in previous studies of the three-body problem \cite{Petrov2004, Gogolin2008, Massignan2008, Schmidt2012, Langmack2018,Chapurin2019, Secker2021, Li2022}.

As was shown in Ref.~\cite{Secker2021_2}, the open-channel projection $\braket*{k, p, \underline{aa}, a}{\bar{\Psi}}$ of the Faddeev component completely decouples from the closed channel $\ket*{\underline{bc},a}$. Since $\braket*{k, p, \underline{aa}, a}{\bar{\Psi}}$ contains the long-range physics associated with the Efimov effect, we only need to solve an effective single channel (ESC) form of Eq. \eqref{eq:STM1}, which contains purely the open-channel component of the two-body transition matrix. Effects of the closed-channel are then included purely at the two-body level. Note that the situation is entirely different for closed channels of the type $\ket*{\underline{ab}}$, where one particle conserves its spin \cite{Secker2021_2}. We choose to exclude this special case from our model, but will comment on it briefly in Sec. \ref{sec:SpinAn}. For closed channels of the type $\ket*{\underline{bb}}$ our model applies without any complications.

\subsection{Effective three-body potential}
\label{sec:pos}

The momentum space approach as outlined in the previous section will be used to solve the three-body problem. However, for the analysis of the Efimovian three-body potential, we need to make a transformation to position space. In this section we give a brief overview of the hyperspherical formalism in which the Efimovian potential is usually expressed. For a more detailed discussion we refer to Refs. \cite{Macek1968,Macek1986, Braaten2006, Nielsen2001}.

For identical particles, the Jacobi dimer separation $r_i$ and atom-dimer separation $\rho_i$ are transformed to a hyperradius $R$ and hyperangle $\alpha_i$,
\begin{align}
\tan{\alpha_i} = \frac{\sqrt{3}r_i}{2\rho_{i}}, \quad R^2 =  r_i^2 + \frac{4}{3}\rho_{i}^2,
\label{eq:HScoordinates}
\end{align}
Here $i=1,2,3$ denotes the Jacobi index, which we will suppress in this section. As the notation implies, $R$ is invariant to a change in Jacobi set. The hyperangle $\alpha$ is often denoted together with the polar and azimuthal angles of the unit vectors $\vu{r}$ and $\vu*{\rho}$ as $\vb*{\Omega}$. We expand the open-channel wave function into a complete and orthonormal set of hyperangular functions $\Phi_{\nu}(R, \vb*{\Omega})$, by the following expansion,
\begin{align}
\begin{split}
\braket*{R, \vb*{\Omega}, \underline{aa}, a}{\Psi} =\frac{1}{R^{\frac{5}{2}}} \sum_{\nu = 0}^{\infty} f_{\nu}(R)\Phi_{\nu}(R, \vb*{\Omega}).
\end{split}
\label{eq:HSexp}
\end{align}
Note that we will suppress spin indices for this expansion as we will always use the open-channel component of the three-body wave function. The hyperangular functions $\Phi_{\nu}(R, \vb*{\Omega})$ are eigenfunctions of the angular momentum part of the three-body Schr\"odinger equation.
The index ${\nu}$ is usually referred to as the hyperspherical channel, and the associated expansion coefficient $f_{\nu}(R)$ as the hyperradial wave function. It obeys the following set of coupled equations \cite{Braaten2003},
\begin{align}
\begin{split}
&\left[-\dv[2]{R} + \frac{\lambda_{\nu}(R) - \frac{1}{4}}{R^2} + \kappa^2 + Q_{\nu\nu}(R) \right] f_{\nu}(R)  \\ & + \sum_{\nu' \neq \nu} \left[Q_{\nu\nu'}(R) + 2P_{\nu\nu'}(R) \dv{R} \right]f_{\nu'}(R) = 0.
\end{split}
\label{eq:hyperrad}
\end{align}
Here $\lambda_{\nu}(R)$ is the eigenvalue associated with the hyperangular function $\Phi_{\nu}(R, \vb{\Omega})$, where $R$ is interpreted as a parameter of the eigenvalue equation. We have also introduced a trimer binding wave number $\kappa = \sqrt{m\abs*{E}/\hbar^2}$, which we use going forward.  Eq. \eqref{eq:hyperrad} defines an infinite set of equations coupled through the presence of the coupling potentials,
\begin{align}
\begin{split}
P_{\nu\nu'}(R) &= - \matrixel*{\Phi_{\nu}}{\pdv{R}}{\Phi_{\nu'}}_{\vb*{\Omega}},\\ & \qquad \text{and} \\ 
Q_{\nu\nu'}(R) &= - \matrixel*{\Phi_{\nu}}{\pdv[2]{R}}{\Phi_{\nu'}}_{\vb*{\Omega}}.
\end{split}
\label{eq:HSCouplePot}
\end{align}
The inner products on the right-hand side should be taken over the space of angular coordinates $\vb*{\Omega}$. The coupling potentials quantify the dependence of the hyperangular distribution on the hyperradius, and are often referred to as non-adiabatic contributions to the three-body problem \cite{Braaten2003}. In the so called scale-free region, where $r_{\mathrm{vdW}} \ll R \ll \abs{a}$, one may show that the eigenvalue $\lambda$ becomes independent of the hyperradius \cite{Braaten2006}. This has the consequence that all non-adiabatic contributions vanish and the coupled set presented in Eq. \eqref{eq:hyperrad} uncouples into single particle Schr\"odinger equations with "effective" hyperradial three-body potentials $U_{m}(R) =  (\lambda_{\nu} - \frac{1}{4})/R^2$. The Efimov channel $\nu=0$ has eigenvalue $\lambda_0 = -s_0^2$, with $s_0 \approx 1.00624$. Thus the associated three-body potential is attractive, inducing the Efimov effect with its characteristic $1/R^2$ scaling. In the scale-free region all channels with $\nu \neq 0$ have associated three-body potentials that are purely repulsive \cite{Braaten2003}.

Due to the non-trivial behavior of the coupling potentials in the short-range regime, the full behavior of the effective potential is very complicated. Upon solving the STM equation \eqref{eq:STM1} however, we can use the three-body wave function to derive an approximation to the Efimovian three-body potential in the scale-free region. First we formulate the three-body probability as,
\begin{align}
\begin{split}
\bar{P}_{\vb*{\Omega}}(R) \sim R^5 \int_{0}^{\frac{\pi}{2}} d\alpha \ \sin^2(2\alpha)\int_{-1}^{1}dx \ \abs{\Psi(R, \alpha, x)}^2,
\end{split}
\label{eq:probint}
\end{align}
where $x = \vu{r} \cdot \vu*{\rho}$ gives the inner product of the two Jacobi vectors, and $\Psi(R, \alpha, x)$ is obtained by a Fourier transformation of Eq. \eqref{eq:FaddeevDecomp}. By virtue of the orthonormality of the hyperangular functions $\Phi(R, \vb*{\Omega})$ it is possible to use the three-body probability to derive an effective three-body potential, a method also applied in Ref. \cite{Naidon2014}. The validity of this method relies on the efficiency of the hyperspherical expansion in Eq. \eqref{eq:HSexp}. Since the Efimov channel is the only attractive channel, we are justified in neglecting all higher lying repulsive channels which suppress the local probability \cite{Wang2012}. Then the expansion contains only one term, and the resulting three-body probability is equal to $\abs*{f_0(R)}^2$. Since we can choose $f_0(R)$ to be real by the normalisation of the wave function, the following expression for an effective three-body potential follows from Eq. \eqref{eq:hyperrad},
\begin{align}
\begin{split}
U_{\mathrm{eff}}(R) = \frac{1}{\sqrt{\bar{P}_{\vb*{\Omega}}(R)}} \dv[2]{R} \sqrt{\bar{P}_{\vb*{\Omega}}(R)} - \kappa^2.
\end{split}
\label{eq:Ueff}
\end{align}
At unitarity this effective potential is expected to be a good approximation of the actual Efimov potential in the scale-free region. In particular it is sufficiently accurate to reproduce the characteristic repulsive barrier around $R \approx 2 \ r_{\mathrm{vdW}}$ and the potential well which appear for broad resonances, as was shown in Ref. \cite{Naidon2014}.

\begin{table}[t]
\caption{Parameters of the physical resonances used as a starting point for our computations. The value of $s_{\mathrm{res}}$ in the last column will be varied with all other parameters held fixed. Data taken from (\textsuperscript{39}K: \cite{Falke2008}), (\textsuperscript{85}Rb: \cite{Claussen2003, Derevianko1999}) and (\textsuperscript{133}Cs: \cite{Chin2010, Berninger2013}).}
\begin{ruledtabular}
\begin{tabular}{ccccc}
Species  & $B_0$ [G]  & $a_{\mathrm{bg}}$ [$r_{\mathrm{vdW}}$] & $\delta \mu$ [$E_{\mathrm{vdW}}/\mathrm{G}$] & $s_{\mathrm{res}}$ \\
\hline
$^{39}\mathrm{K}$  & 33.50 & -0.31 & -0.154 & 2.46 \\
$^{85}\mathrm{Rb}$  & 155.04  & -5.40 & -0.517 & 28.6  \\
$^{133}\mathrm{Cs}$  & -11.7  & 17.02 & 1.21  & 565 \\
$^{133}\mathrm{Cs}$ & 547.0  & 24.74 & 0.94 & 167 \\
\end{tabular}
\end{ruledtabular}
\label{tab:resparam}
\end{table}

\section{Results}
\label{sec:res}

To fix the degrees of freedom in the model of Sec. \ref{sec:2chan} we take sets of resonance parameters measured from physical resonances, summarised in Tab. \ref{tab:resparam}. We then shift the resonance strength $s_{\mathrm{res}}$ away from the physical value by altering $\Delta B$, keeping all other parameters fixed. As reported in Ref. \cite{Langmack2018}, the change in the three-body parameter with varying resonance strength becomes more abrupt as $a_{\mathrm{bg}}$ approaches the value of $a_-$ in the open-channel potential.
We check whether our model reproduces this behavior by artifically altering the Rubidium resonance in Tab. \ref{tab:resparam} such that $a_{\mathrm{bg}} = - 9.75 \ r_{\mathrm{vdW}}$, noting that the open-channel potential has a three-body parameter $a_- = -10.85 \ r_{\mathrm{vdW}}$ in the EST approximation.

\subsection{Efimov spectra}
\label{sec:spec}
\begin{figure}[t]
\centering
\includegraphics[height=0.35\textwidth]{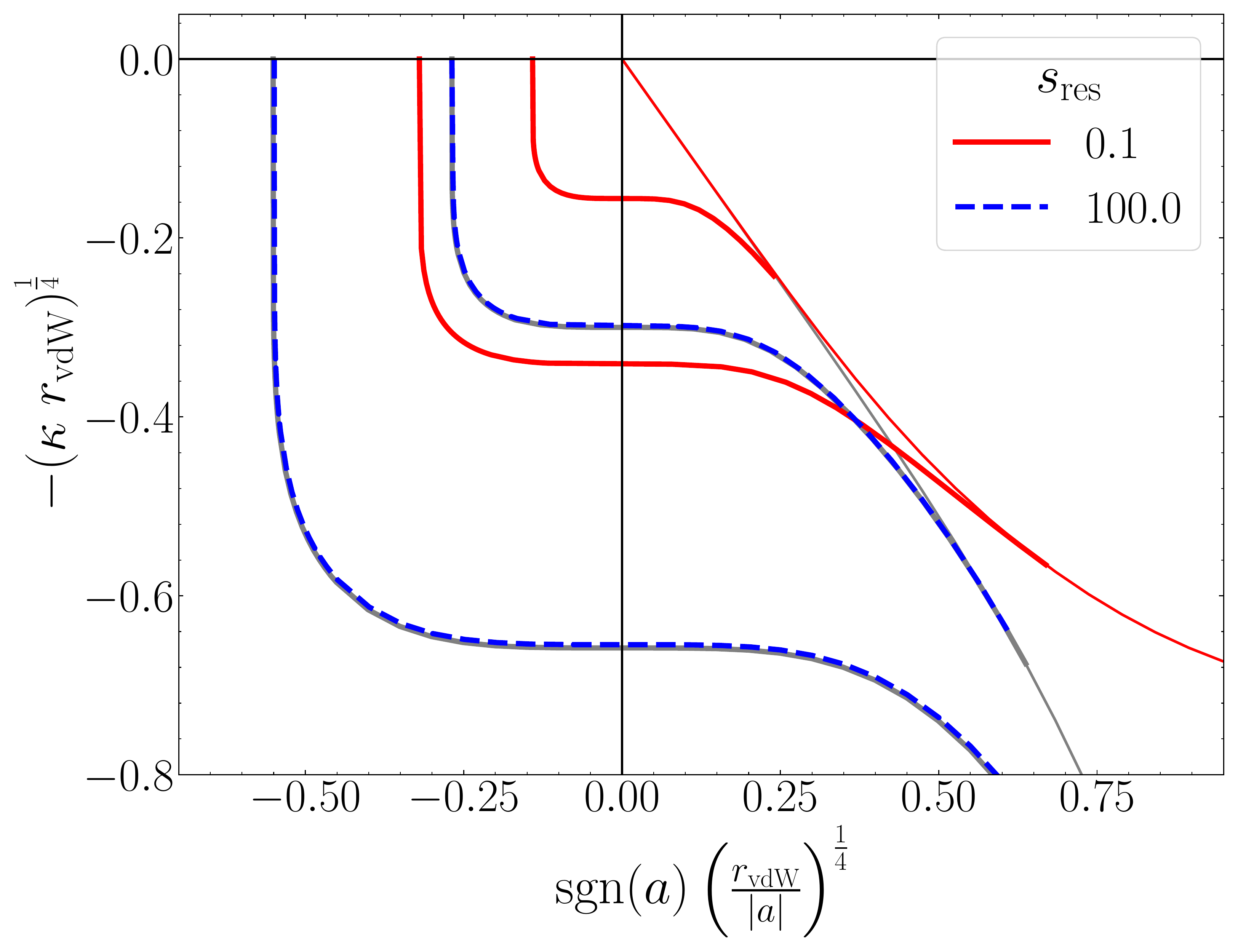}
\caption{Binding wave number of the two lowest lying Efimov states as a function of the inverse scattering length, computed at two different resonance strengths. For $a>0$ the atom-dimer thresholds from Fig.~\ref{fig:EfPlot_dim} are plotted with thin lines. Grey lines show the energies obtained from a single-channel EST model, which corresponds with the broad-resonance limit $s_{\mathrm{res}}\rightarrow \infty$. Multichannel model used to produce this figure was tuned to the \textsuperscript{39}K resonance in table \ref{tab:resparam}.}
\label{fig:EfPlot_sr}
\end{figure}

We first apply our model to the computation of the Efimov spectra and associated three-body parameters as a function of the Feshbach resonance strength. First, we show in Fig.~\ref{fig:EfPlot_sr} the spectrum of the two lowest lying Efimov states for finite scattering lengths surrounding the Feshbach resonance. Comparing with the broad resonance limit $s_{\mathrm{res}} \rightarrow \infty$, we find that as the resonance strength is decreased the Efimov spectrum is squeezed into a smaller area of the $(\kappa, 1/a)$ plane, corresponding to an increase of the three-body parameter $\abs{a_-}$. Alternatively one can also define the three-body parameter via the wavenumber $\kappa_*$ of the ground state trimer at resonance, which decreases as the resonance becomes narrow.

\begin{figure*}
     \centering
               \subfloat[\label{fig:ksam_1}]{
\includegraphics[width = 0.49\textwidth]{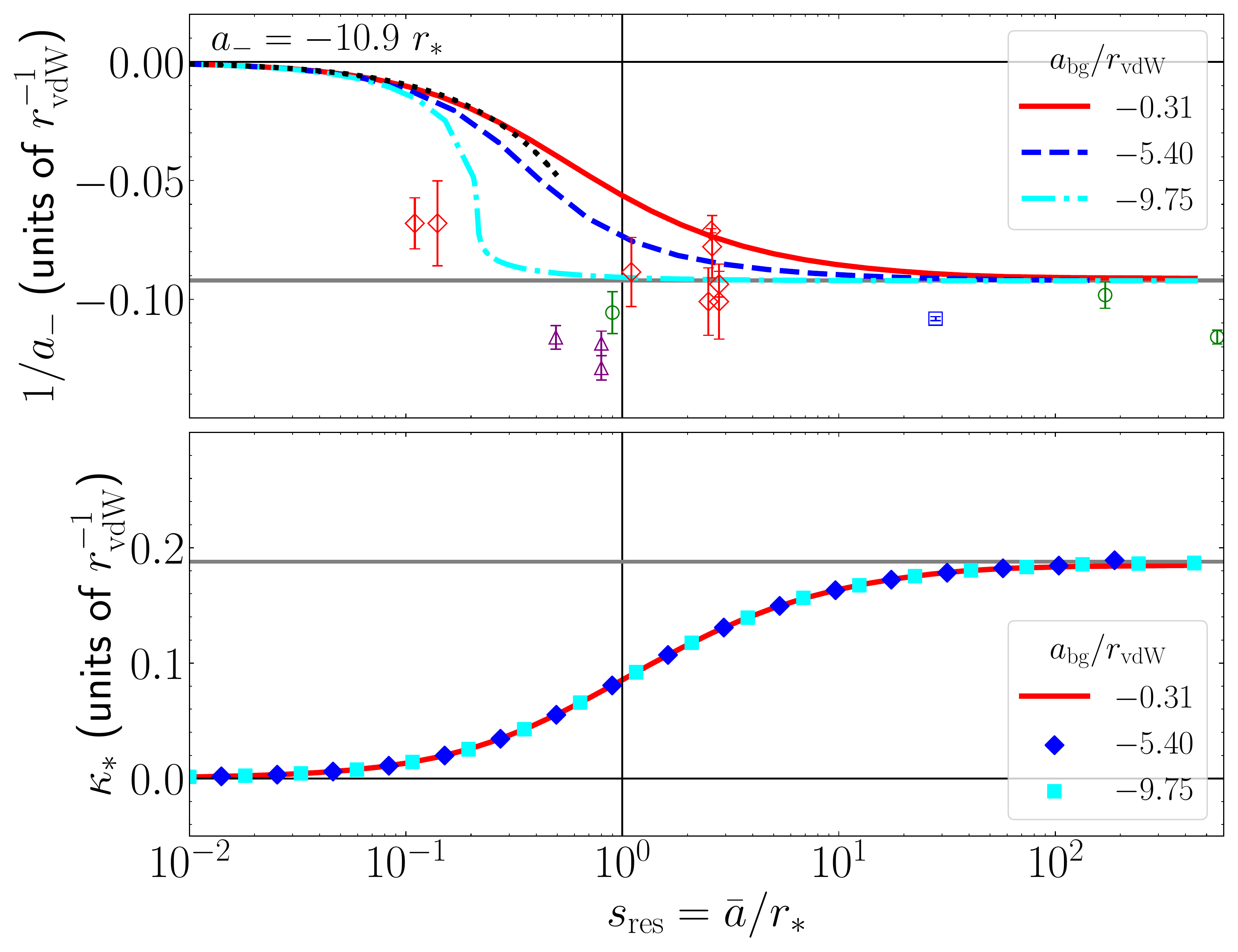}}
               \subfloat[\label{fig:ksam_2}]{
\includegraphics[width = 0.49\textwidth]{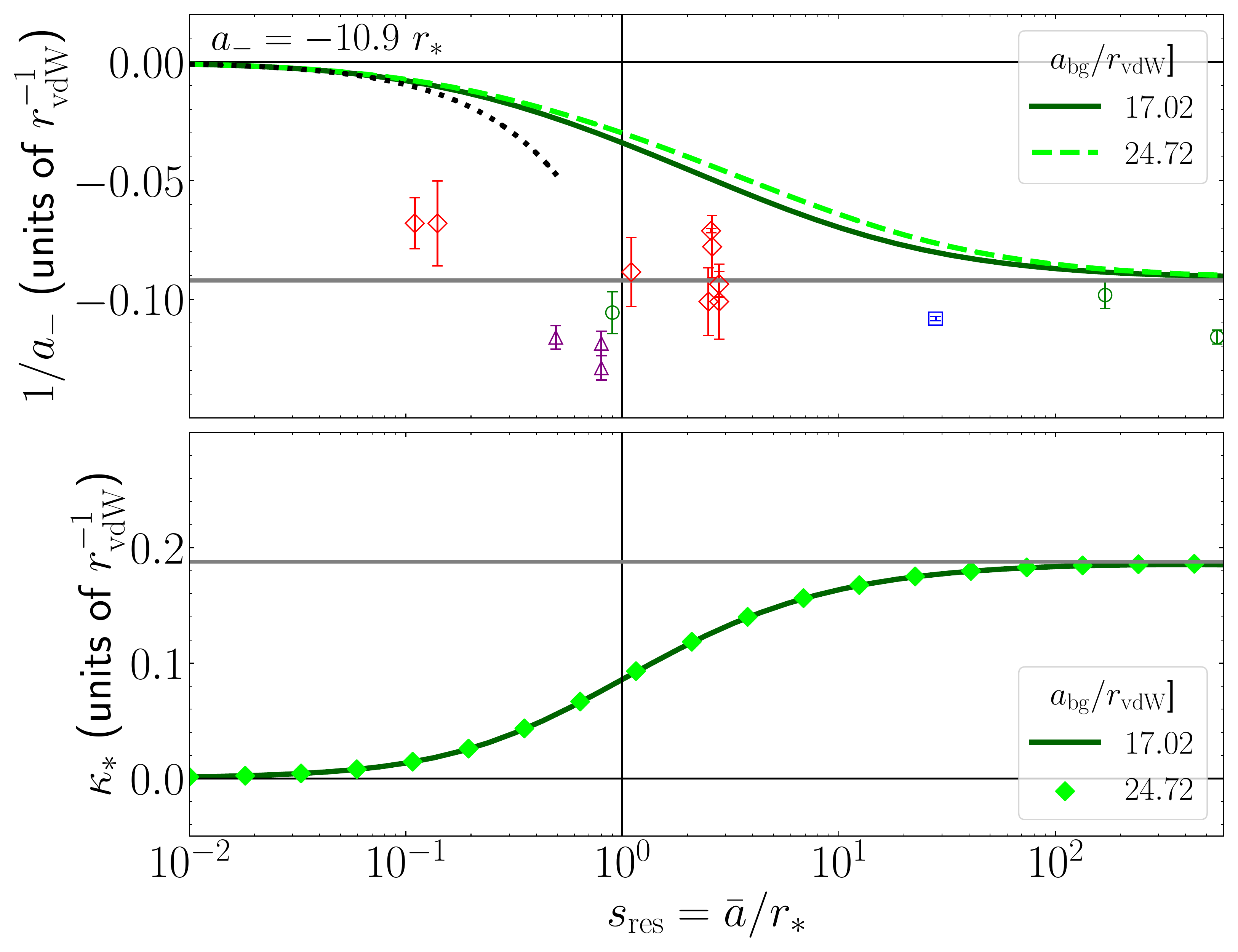}}
     \caption{Plots of the three-body parameters as a function of resonance strength, for negative $a_{\mathrm{bg}}$ in Fig. (a) and positive $a_{\mathrm{bg}}$ in Fig. (b). In the top panels we show the inverse of the dissociation scattering length $a_-$ associated with the ground state Efimov trimer. Here colored lines show the results of our multichannel model, for different background scattering lengths. Scatter points represent experimental data, measured in $^{133}\mathrm{Cs}$ (green circles) \cite{Berninger2011}, $^{7}\mathrm{Li}$ (purple triangles) \cite{Gross2011, Dyke2013}, $^{85}\mathrm{Rb}$ (blue squares)  \cite{Wild2012} and $^{39}\mathrm{K}$ (red diamonds) \cite{Roy2014,Chapurin2019}. The color of the line plots is chosen to match the data points, e.g. the red plots were computed using resonance parameters matching a resonance in $^{39}\mathrm{K}$.  In the narrow resonance limit $s_{\mathrm{res}} \ll 1$ we illustrate the limiting behavior $a_- = -10.9 \ r_*$ as a black dotted line. In the bottom panels we plot the trimer binding wavenumber $\kappa_*$ at resonance. Here, results for different background scattering length practically overlap on the scale shown, so we use scatter plots to distinguish between them. In both the upper and lower panels the three-body parameter obtained with a single-channel EST model, correspondent with the broad resonance limit $s_{\mathrm{res}} \gg 1$, is shown as a grey horizontal line.}
\label{fig:ksam}
\end{figure*}

The shift in three-body parameters can be more clearly seen in Fig.~\ref{fig:ksam}, where we show a scan of $\abs*{a_-}$ and $\kappa_*$ from the broad to narrow resonance limit. For the sake of comparison, Fig.~\ref{fig:ksam} also contains experimental data for a select set of physical resonances. As the resonance strength decreases, the three-body parameter $\abs{a_-}$ shows a monotonous increase, consistent with findings in earlier studies such as Refs. \cite{Schmidt2012, Langmack2018}. Within the narrow resonance limit our results approach the universal limit $a_- = - 10.9 \ r_*$. The effect of the background scattering length can mainly be observed in the intermediate strength regime, where larger negative values of $a_{\mathrm{bg}}$ tend to push the three-body parameter closer to the universal broad resonance value for a larger portion of the resonance strength regime. This finding is consistent with more artificial models of the two-body interaction, such as the approach adopted in Ref. \cite{Langmack2018}. However, our more realistic EST model strongly suppresses the sensitivity of the three-body parameter to the background scattering length. Similarly, we find that replacing our EST potential by a simple ultraviolet cut-off greatly increases the sensitivity to $a_{\mathrm{bg}}$. This is especially true for the binding wave number $\kappa_*$, whose dependence on $a_{\mathrm{bg}}$ is negligible on the scale of Fig.~\ref{fig:ksam}. In Ref. \cite{Langmack2018}, sensitivity of $\kappa_*$ to $a_{\mathrm{bg}}$ was attributed to beyond effective range effects. For van der Waals potentials, the effective range $r_e^{\mathrm{vdW}}$ as a function of $s_{\mathrm{res}}$ is known to behave as \cite{Gao2011, Werner2012},
\begin{align}
\begin{split}
r_e^{\mathrm{vdW}} &= \frac{\bar{a}}{6} \left(\frac{\Gamma\left(\frac{1}{4} \right)}{\Gamma\left(\frac{3}{4} \right)} \right)^2 \left[1 - 2\frac{\bar{a}}{a} + 2\left(\frac{\bar{a}}{a}\right)^2 \right]\\& -\frac{\bar{a}}{s_{\mathrm{res}}} \left[ 1 - \frac{a_{\mathrm{bg}}}{a}\right]^2 
\end{split}
\end{align}
The first term contains all non-resonant contributions that give a universally determined effective range at unitarity \cite{Flambaum1999}, while the second term captures the additional contribution due to the Feshbach resonance. Since the EST model reproduces the zero-energy wave function exactly, the effective range in our multichannel model is well described by this equation. At resonance, where $a_{\mathrm{bg}}/a \rightarrow 0$, the effective range becomes fully independent of the background scattering length similar to the three-body parameter $\kappa_*$. 

\subsection{Three-body repulsion}
\label{sec:repuls}

\begin{figure}[t]
\centering
\includegraphics[height=0.35\textwidth]{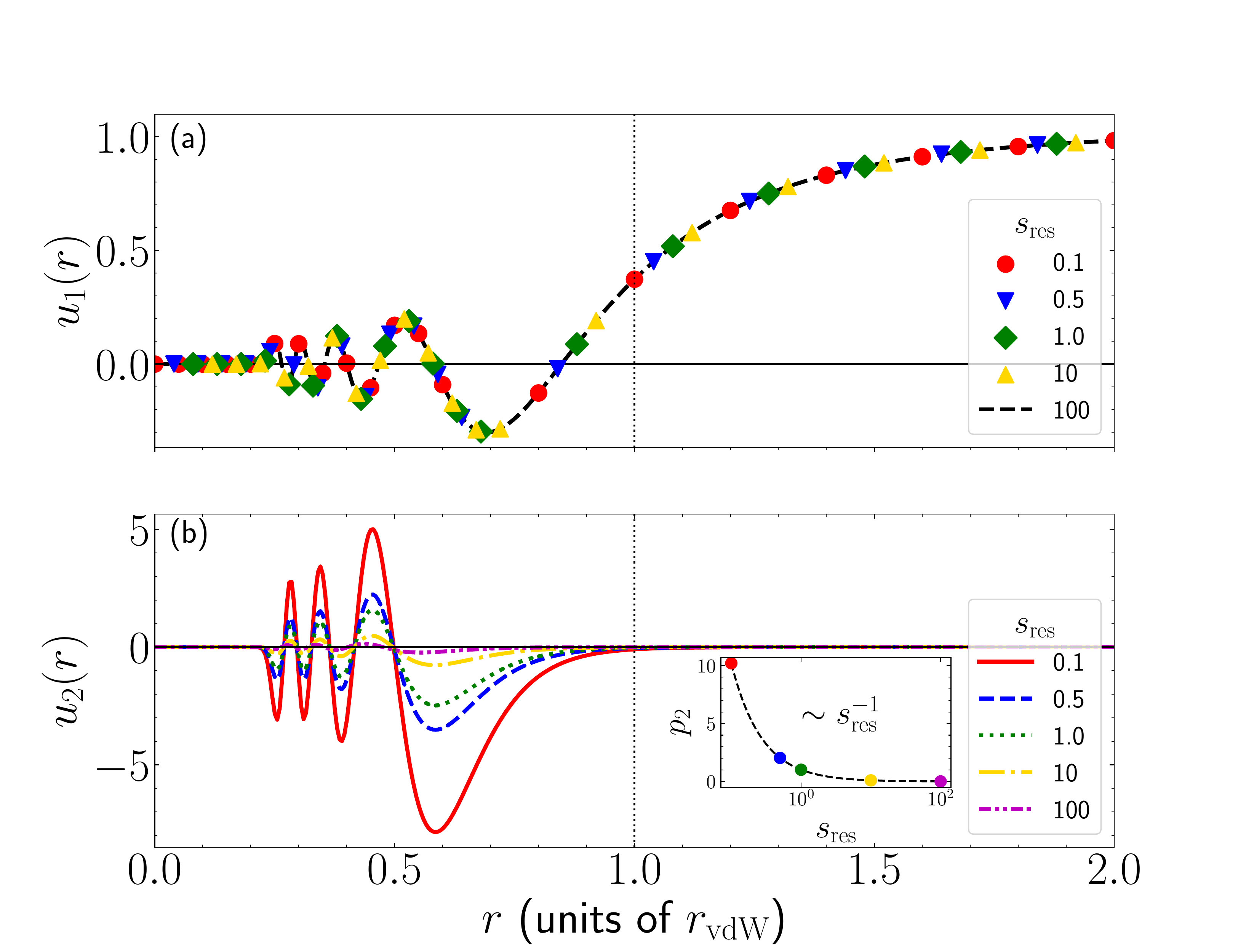}
\caption{Open (a) and closed (b) channel functions $u_{1}(r)$ and $u_{2}(r)$ for different values of the resonance strength, computed at unitarity for the Feshbach resonance with $a_{\mathrm{bg}} = -0.31 \ r_{\mathrm{vdW}}$. The $r= r_{\mathrm{vdW}}$ boundary is emphasized by the black dotted line. The normalisation is chosen such that the total wave function asymptotes to unity in the long range. The inset in figure (b) shows the scaling of the integrated closed-channel two-body probability $p_2 = \int dr \left| u_{2}(r) \right|^2$ with $s_{\mathrm{res}}$.}
\label{fig:2body_sr}
\end{figure}

Having confirmed that the three-body parameter in our model scales as expected in both the narrow and broad resonance limits, we now move on to a position space analysis using the formalism of Sec. \ref{sec:pos}. Given that our results are largely insensitive to $a_{\mathrm{bg}}$ at the position of the resonance, we limit ourselves in this section to the \textsuperscript{39}K resonance with $a_{\mathrm{bg}} = -0.31 \ r_{\mathrm{vdW}}$. Before analyzing the three-body state directly, it is instructive to consider the two-body scattering wave function $\braket*{r}{\psi}$ (see Eq. \eqref{eq:2bpsi}) that is used to construct the EST potential. To this end we plot the open and closed channel radial components of the wave function for a set of different resonance strengths in Fig.~\ref{fig:2body_sr}, normalised such that the wave function asymptotes to $1$ for $r \gg r_{\mathrm{vdW}}$. As predicted by multichannel resonance theory the open-channel amplitude $\abs{u_1(r)}$ is independent of $s_{\mathrm{res}}$, whilst the closed-channel amplitude $\abs{u_2(r)}$ scales as $\sim 1/\sqrt{s_{\mathrm{res}}}$ \cite{Kohler2006}. Previous analyses of single-channel van der Waals interactions have connected the suppression of the two-body wave function below distances of $1 \ r_{\mathrm{vdW}}$ to the appearance of a universal repulsive barrier in the three-body potential at $R \approx 2 \ r_{\mathrm{vdW}}$ \cite{Wang2012, Naidon2014}. In the two-channel case this suppression persists in the open-channel component, whilst there appears an increase in total short-range two-body probability due to finite lifetime of the closed-channel state.

\begin{figure}
\centering
\includegraphics[height=0.35\textwidth]{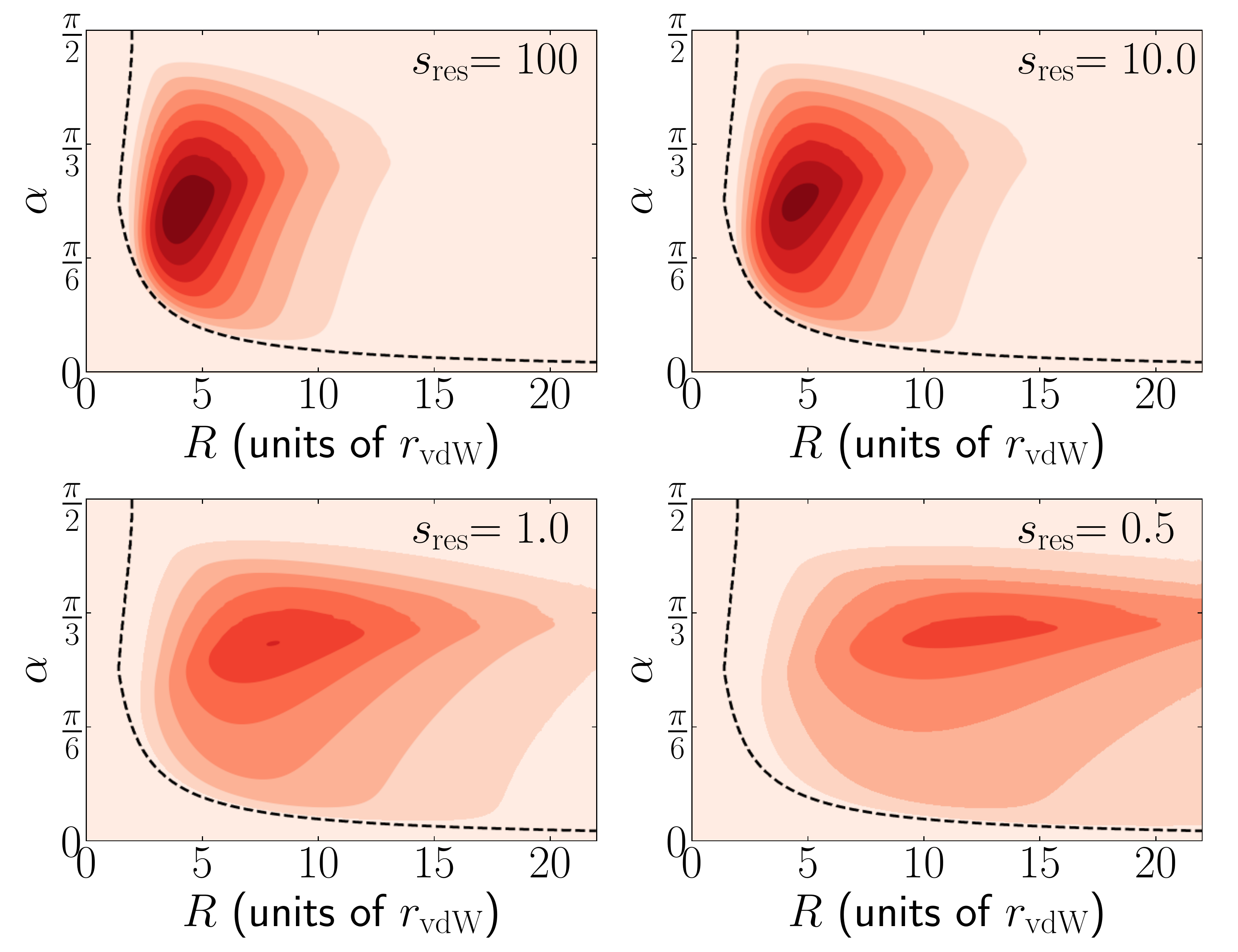}
\caption{Contour plots of the three-body probability at four different values of $s_{\mathrm{res}}$. Computed using the $^{39}\mathrm{K}$  resonance with $a_{\mathrm{bg}} = -0.31 \ r_{\mathrm{vdW}}$ as input. The boundary below which any two particles approach closer than $1 \ r_{\mathrm{vdW}}$ is shown by a black dashed line. Format inspired by Ref. \cite{Naidon2014}}
\label{fig:cont_sr}
\end{figure}

To examine the effect of the closed-channel on the three-body level we first compute the three-body probability in the $\left\{R, \alpha \right\}$ plane, by omitting the integration over the hyperangle in Eq. \eqref{eq:probint}. The results are shown in Fig.~\ref{fig:cont_sr}. Here the effect of the open-channel suppression is highlighted by plotting the boundary beyond which any two particles approach below the van der Waals length. As expected, the three-body probability is strongly suppressed beyond this boundary for the broad resonance (first two panels of Fig.~\ref{fig:cont_sr}), where the closed-channel component is small. Interestingly, as we tune our interaction towards the narrow resonance regime and the closed-channel amplitude increases, we see no additional penetration of the region of open-channel suppression. Instead, we find that both the average and the spread of the three-body wave function in the hyperradial coordinate increase. This suggests that the open-channel suppression of two-body probability remains a dominant factor for small hyperradii, strongly suppressing the three-body probability regardless of resonance strength. The increase in closed-channel two-body amplitude mainly impacts the intermediate to long distance regime where $R > 2 \ r_{\mathrm{vdW}}$. As has been noted before \cite{Wang2012, Naidon2014}, the appearance of a three-body repulsive barrier is associated with a repulsive potential energy peak due to the non-adiabatic correction $Q_{00}(R)$ in Eq. \eqref{eq:hyperrad}, reminiscent of an angular momentum barrier. This peak arises due to a squeezing of the hyperangular distribution function $\Phi(R,\vb{\Omega})$ as $R$ decreases, driven by the short-range two-body suppression. Our results in Fig.~\ref{fig:cont_sr} show that the location of this barrier remains universally determined by the van der Waals length also near a narrow resonance.

\begin{figure}
\centering
\includegraphics[height=0.4\textwidth]{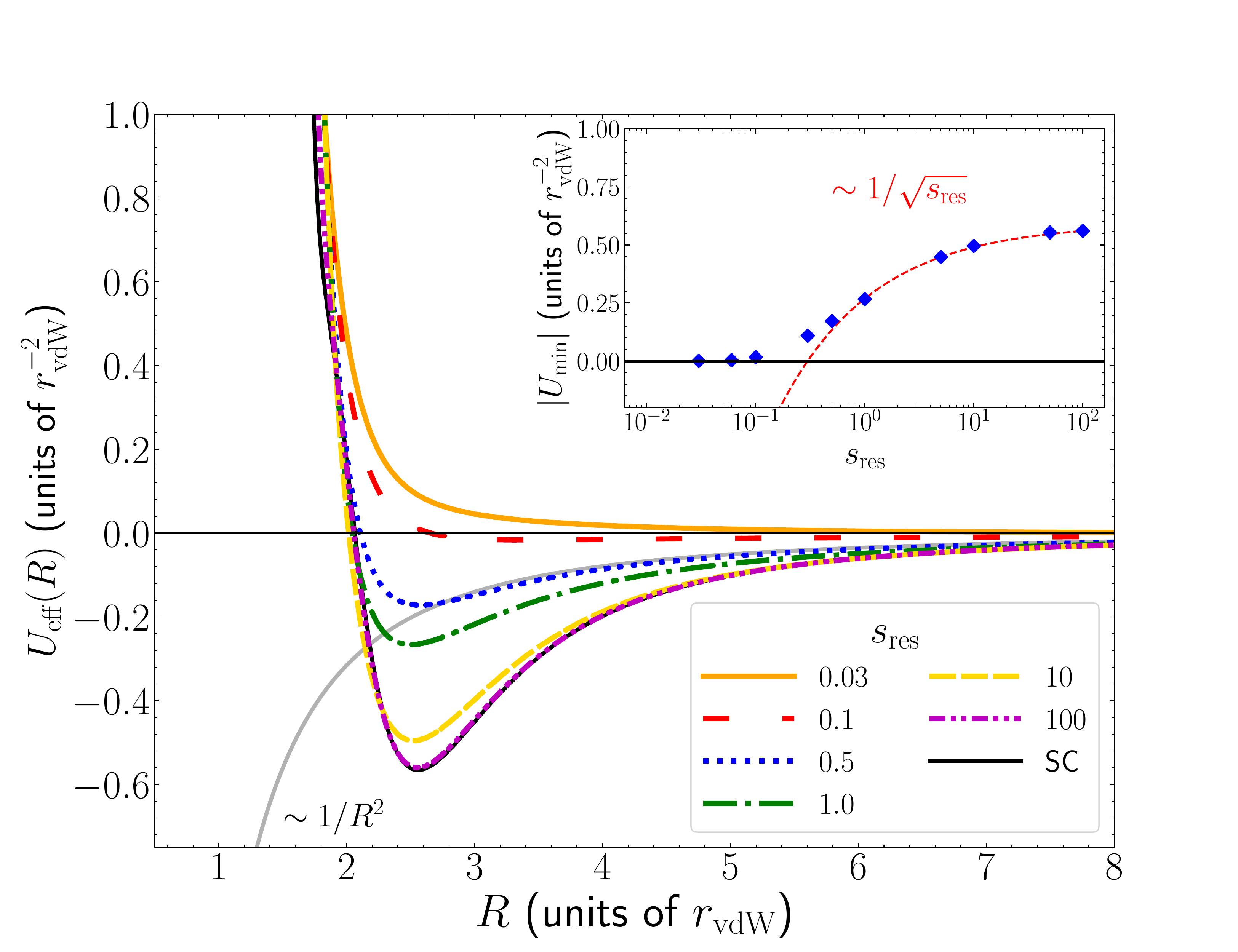}
\caption{Plot of the effective three-body potential as a function of the hyperradius, computed with starting parameters taken from the $^{39}\mathrm{K}$  resonance as outlined in table \ref{tab:resparam}. Colored lines show the effective potential at different values of the resonance strength parameter. In the regime of small hyperradius $R \lesssim r_{\mathrm{vdW}}$ there appear unphysical and meaningless oscillations in the potential as artifacts of the model \cite{Naidon2014}, which we have removed from the plot to avoid unneccessary clutter. Black line shows the effective potential computed by using the associated single-channel EST model. Grey line shows the asymptotic $\sim 1/R^2$ Efimov attraction as follows from the zero-range theory. Plot includes an inset showing the scaling of the minimum $U_{\mathrm{min}}(R)$ of the effective potential with $s_{\mathrm{res}}$.}
\label{fig:EfPot_sr}
\end{figure}

We now proceed by integrating out the hyperangle to obtain $\bar{P}_{\vb*{\Omega}}(R)$ and use Eq. \eqref{eq:Ueff} to derive the effective three-body potential. The results are plotted in Fig.~\ref{fig:EfPot_sr}, where for the sake of comparison we also show the result with a single-channel interaction (correspondent with $s_{\mathrm{res}} \rightarrow \infty$), and the universal $\sim1/R^2$ potential from zero-range theory which corresponds with the limit $R/r_{\mathrm{vdW}} \rightarrow \infty$ \cite{Braaten2006}. Consistent with Fig.~\ref{fig:cont_sr}, a decreasing resonance strength manifests most strongly in the intermediate to long distance regime, where we observe a strong decrease in the depth of the effective potential that pushes the Efimov state closer to threshold. This corresponds with a decrease of the binding energy $\kappa_*$, as observed in Fig.~\ref{fig:ksam}. To verify whether this behavior continues into the narrow resonance limit, we have tracked the effective potential up to $s_{\mathrm{res}} = 0.01$. Here the potential at larger separations becomes practically flat, signifying that all hyperradii have approximately equal probability. To obtain a more quantitative characterisation of the decrease in depth we plot the minimum of the effective potential as a function of $s_{\mathrm{res}}$, shown in the inset of Fig.~\ref{fig:EfPot_sr}. In the broad resonance limit we find that the depth of the barrier scales with $1/\sqrt{s_{\mathrm{res}}}$, and is hence inversely proportional to the closed-channel amplitude as plotted in Fig.~\ref{fig:2body_sr}. Consistent with Fig.~\ref{fig:cont_sr} the position of the repulsive barrier is set by the van der Waals length with the relation $R \approx 2 \ r_{\mathrm{vdW}}$, regardless of the resonance strength. 


To supplement our findings we have also computed three-body effective potentials for particles interacting via exponentially decaying potentials, common in nuclear physics. Here the three-body parameter has been predicted to be likewise different compared to the single-channel van der Waals interaction \cite{Naidon2014_2}. However, in contrast to the multichannel case we find that this change originates from a shift in the location of the three-body repulsive barrier, see Appendix \ref{ap:short} for more detail. The contrast highlights the underlying multichannel nature in our observation in Fig.~\ref{fig:EfPot_sr}.

\subsection{Analysis in spin-position space}
\label{sec:SpinAn}
To gain a better physical understanding of the origin of the observations made in the previous section, it is instructive to consider explicitly the multichannel structure of the three-body problem. In the open-channel state $\ket*{\underline{aa},a}$, every pair of particles interacts via the strong resonant van der Waals interaction which induces the Efimov effect. Together with the suppression of two-body probability when $r < r_{\mathrm{vdW}}$, this will drive the particles towards equilateral three-body configurations which minimize the likelihood of small nuclear distances. As mentioned in the previous section, these dynamics are recognized in the hyperspherical picture via the non-adiabatic potential $Q_{00}(R)$, which forms a strong repulsive barrier at small hyperradii \cite{Wang2012, Naidon2014}. In the multichannel case, the physical picture is complicated by the presence of a three-body closed channel $\ket*{\underline{bc},a}$. In this state there appears an asymmetry in the strength of the two-particle interactions, given that two of the three pairs exist in the non-resonant channels $\ket*{\underline{ab}}$ and $\ket*{\underline{ac}}$. Hence the interaction with the third particle is much weaker than the interaction felt by two particles in the $\ket*{\underline{aa}}$ state, which is resonantly enhanced.

\begin{figure}[t]
\centering
\includegraphics[height=0.25\textwidth]{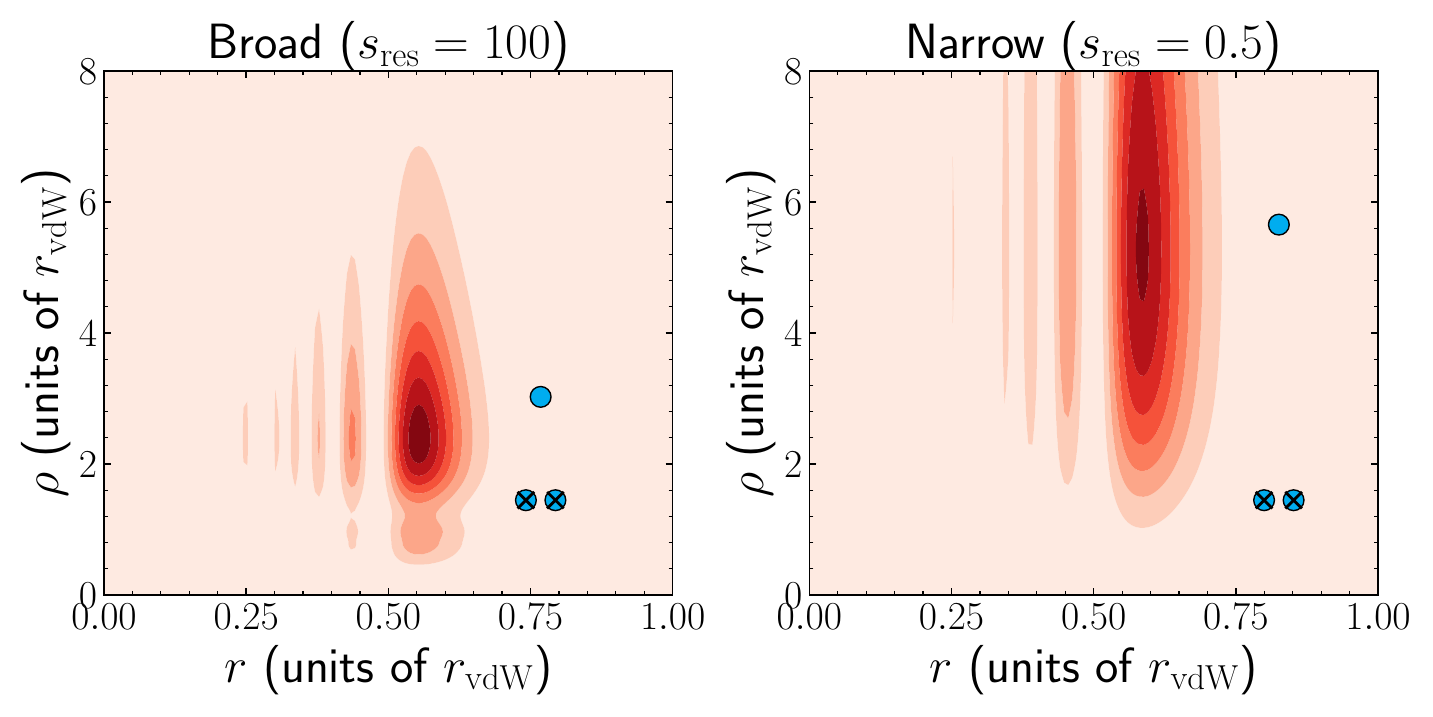}
\caption{Contour plots of three-body probability $P_{\underline{bc},a}(r, \rho) \sim r^2 \rho^2 \abs*{\braket*{r\rho;\underline{bc},a}{\Psi}}^2$ in the plane of Jacobi coordinates $r$ and $\rho$, for a broad and narrow Feshbach resonance. Note that $\braket*{\underline{bc},a}{\Psi} = \braket*{\underline{bc},a}{\bar{\Psi}}$ since $\braket*{\underline{ab},c}{\bar{\Psi}}$ and $\braket*{\underline{ac},b}{\bar{\Psi}}$ vanish by neglecting non-resonant interactions \cite{Secker2021_2}.  The drawings show the changing structure of the Efimov state, where particles in the closed channel $\ket*{\underline{bc}}$ are drawn with a cross.}
\label{fig:Psi_bba}
\end{figure}

With these effects in mind we now turn our attention once more to the results presented in Fig.~\ref{fig:EfPot_sr}. As we decrease $s_{\mathrm{res}}$, the lifetime of the closed-channel state increases, scaling as $1/s_{\mathrm{res}}$. The third atom then interacts with the pair $\ket*{\underline{bc}}$ via a non-resonant van der Waals interaction, much weaker than the resonant interaction present for the broad resonance. This leads to a gradual decrease of the depth of the effective potential shown in Fig.~\ref{fig:EfPot_sr}, as the coupling to the closed channel $\ket*{\underline{bc},a}$ stretches the three-body state to more elongated configurations. To illustrate this behavior we have computed the closed-channel component $\braket*{\underline{bc},a}{\bar{\Psi}}$ of the three-body wave function, which can be directly obtained from Eq. \eqref{eq:FadComp} once $F_a(p)$ has been computed. We then formulate the closed-channel three-body probability $P_{\underline{bc},a}$, plotted in Fig.~\ref{fig:Psi_bba}. Here one clearly observes the stretching of the wave function that occurs near a narrow resonance, which is directed along the $\rho$ coordinate quantifying the separation of the third particle. In the limit of a very narrow resonance, the third particle is free to drift towards separations far beyond $r_{\mathrm{vdW}}$, consistent with a flat three-body potential. In contrast, the shape of the probability along the dimer separation $r$ is relatively unaffected by the resonance strength, and in fact follows the structure of the two-body closed channel wave function as shown in Fig.~\ref{fig:2body_sr}. Fig.~\ref{fig:Psi_bba} also shows that there is no repulsive barrier at small hyperradii in the closed channel state, which allows particles to approach to within the van der Waals range. The fact that a universal short-range repulsive barrier remains in $\braket*{\underline{aa},a}{\bar{\Psi}}$ also for narrow resonances, as shown in Figs.~\ref{fig:cont_sr} and \ref{fig:EfPot_sr}, is due to the influence of the open-channel component, where a universal barrier due to $Q_{00}(R)$ always exists and is independent of resonance strength. This prevents coupling to the closed-channel state for small hyperradii, hence preserving the short-range suppression of the wave function, as clearly observed in Figs.~\ref{fig:cont_sr} and \ref{fig:EfPot_sr}.

As was discussed in Sec. \ref{sec:ESC_EST}, the relative weakness of non-resonant interactions is a commonly used assumption for the three-body problem. It is important to note that while this assumption is valid for most systems, there are special cases where it is expected to be incorrect. For example, if the resonant closed channel state is taken as $\ket*{\underline{ab}}$, i.e. just one particle changes its state, then both the open and closed channel three-body states have purely resonant interactions. We expect this to alter the behavior of the three-body potential, and indeed it was shown in Ref. \cite{Secker2021_2} that for closed channels of this type the scaling of the three-body parameter with $s_{\mathrm{res}}$ is actually inverted. Additionally, our model is not expected to hold for strongly overlapping Feshbach resonances, where several channels have resonantly enhanced interactions simultaneously.

\section{Conclusion and outlook}
\label{sec:conc}

In this work we have analysed the change in the Efimovian three-body potential as the Feshbach resonance strength is tuned from the broad to narrow resonance regime. For this purpose we have developed a two-channel separable model that takes into account the full coupled-channels low-energy scattering wave function. Our numerical results show that as the resonance strength is tuned away from the broad limit, the associated change in the three-body parameter $a_-$ originates from a decrease of the three-body potential depth in the intermediate distance regime where $R > 2 \ r_{\mathrm{vdW}}$. In contrast, the three-body repulsive barrier that is observed in single-channel models at $R \approx 2 \ r_{\mathrm{vdW}}$ remains universally determined by the van der Waals length. We have interpreted our observations to originate from the relative weakness of interactions between non-resonant spin-channels compared to the resonant interaction that exists in the open channel and drives the Efimov effect. Hence our results should apply generally to systems in which the Feshbach resonance is sufficiently isolated.

There are several possible opportunities for extensions of our approach. Our physical picture of the decreasing three-body attraction for narrow resonances rests on the assumption that the interaction between closed and open-channel particles is off-resonant, such that it may be neglected. Consequently we expect that the presence of a resonant third-channel alters the behavior of the potential significantly, which could be accurately captured in a three-channel EST model. Another point of interest is the analysis of special closed-channel configurations of the type $\ket*{\underline{ab}}$, where it is known that the value of $\abs*{a_-}$ decreases for a narrow resonance \cite{Secker2021_2}. Such a system however is not easily analysed with our model, since the trimer wave function becomes more localised in the short-range where the effective three-body potential is not a useful construct.

\begin{acknowledgments}

We thank Pascal Naidon and Thomas Secker for discussions. This research is financially supported by the Dutch Ministry of Economic Affairs and Climate Policy (EZK), as part of the Quantum Delta NL programme, and by the Netherlands Organisation for Scientific Research (NWO) under Grant No. 680-47-623.

\end{acknowledgments}

\appendix

\section{Tuning of Feshbach resonance parameters}
\label{ap:2bodytuning}
In this appendix we given some additional detail on the tuning of the parameters of our two-body multichannel interaction. A Feshbach resonance is typically parametrised by the following relation \cite{Chin2010},
\begin{align}
\begin{split}
a(B) = a_{\mathrm{bg}} \left(1 - \frac{\Delta B}{B - B_0} \right).
\end{split}
\label{eq:fesa}
\end{align}
The background scattering length $a_{\mathrm{bg}}$ can be set directly by tuning the short-range parameter $r_0$. The resonance width $\Delta B$, is given as \cite{Chin2010},
\begin{align}
\Delta B \underset{k \rightarrow 0}{=} \frac{\pi}{a_{\mathrm{bg}}k\delta \mu} \abs{\matrixel*{\phi_{\mathrm{res}}}{V_{21}}{\psi_{\varepsilon}}}^2.
\label{eq:reswidth}
\end{align}
Here $\ket*{\phi_{\mathrm{res}}}$ is the unit normalised wave function of the resonant bound state, $\ket*{\psi_{\varepsilon}}$ is the energy normalised scattering wave function in the open-channel, and $k = \sqrt{m \varepsilon/\hbar^2}$. We vary the parameters $\alpha$ and $\beta$ until Eq. \eqref{eq:reswidth} is satisfied for a given $\Delta B$. Note that this mapping is not unique, but we have numerically confirmed our results to be insensitive to different choices of $\alpha$ and $\beta$. The last parameter to fix is the bare resonance position $B_{\mathrm{res}}$, which is shifted to $B_0$ by the presence of the spin-exchange interaction. In van der Waals potentials it is possible to approximate the relation between $B_0$ and $B_{\mathrm{res}}$ using the techniques of multichannel quantum defect theory (MQDT) \cite{Greene1982, Mies1984, Mies1984_2, Jachymski2013, Naidon2019}. This leads to the direct relation,
\begin{align}
\begin{split}
B_0 = B_{\mathrm{res}}+\left[\frac{r_{\mathrm{bg}}(1 - r_{\mathrm{bg}})}{1+(1-r_{\mathrm{bg}})^2} \right] \Delta B,
\end{split}
\label{eq:2chanmodshift}
\end{align}
where $r_{\mathrm{bg}} = a_{\mathrm{bg}}/ \bar{a}$. With Eqs. \eqref{eq:fesa}, \eqref{eq:reswidth} and \eqref{eq:2chanmodshift} we can fix all the parameters of our model.

\section{EST two-body transition matrix}
\label{ap:EST}
In this appendix we give explicit expressions for the momentum projection of the separable transition matrix in our multichannel EST separable potential. 
The function $\tau(z)$ follows from Eq. \eqref{eq:LSt},
\begin{align}
\begin{split}
&\tau^{-1}(z) =  \frac{m}{\hbar^2} \bigg[\frac{2 \pi^2}{a} \abs{g_{1}(0)}^2 \\ & +  4\pi \sum_{\sigma} \int_0^{\infty} dk \frac{k^2 \left(\frac{mz}{\hbar^2} \right) \abs{g_{\sigma}(k)}^2}{\left(k^2 + \frac{m \varepsilon_{\sigma}}{\hbar^2} \right) \left(k^2 + \frac{m \varepsilon_{\sigma}}{\hbar^2} - \frac{m z}{\hbar^2} \right)} \bigg],
\end{split}
\label{eq:tau}
\end{align}
This form is inspired by Ref. \cite{Naidon2017}, and uses the fact that the zero-energy on-shell transition matrix is related to the scattering length as $t_{1,1}(0,0,0) = \hbar^2 a /(2\pi^2m)$. The form factors can be computed directly from Eq. \eqref{eq:ESTdef}, by inserting complete sets of position states. Our normalisation gives $\braket*{\vb{r}}{k,\sigma} \sim \sin{(kr)}/(kr) \ket*{\sigma}$, and we expand the scattering wave function into channel functions $u_{\sigma}(r)$ as,
\begin{align}
\begin{split}
\braket*{\vb{r}}{\psi} \sim \sum_{\sigma} \frac{u_{\sigma}(r)}{r} \ket{\sigma}.
\end{split}
\label{eq:2bpsi}
\end{align}
Then the form factors, normalised such that $g_{1}(0) = 1$, are given by,
\begin{align}
\begin{split}
g_{\sigma}(k) = \frac{\sum_{\sigma'} \int_0^{\infty} dr \ \sin{\left(kr\right)} \  V_{\sigma ,\sigma'}(r) \ u_{\sigma'}(r)}{ k \sum_{\sigma'} \int_0^{\infty} dr \ r \ V_{1,\sigma'}(r) \ u_{\sigma'}(r)},
\end{split}
\label{eq:FormFac}
\end{align}
where $V_{\sigma, \sigma'} = \matrixel*{\sigma}{V}{\sigma'}$.

\section{Comparison of resonant to non-resonant interactions}
\label{ap:comp}

\begin{figure}
\centering
\includegraphics[width = 0.45\textwidth]{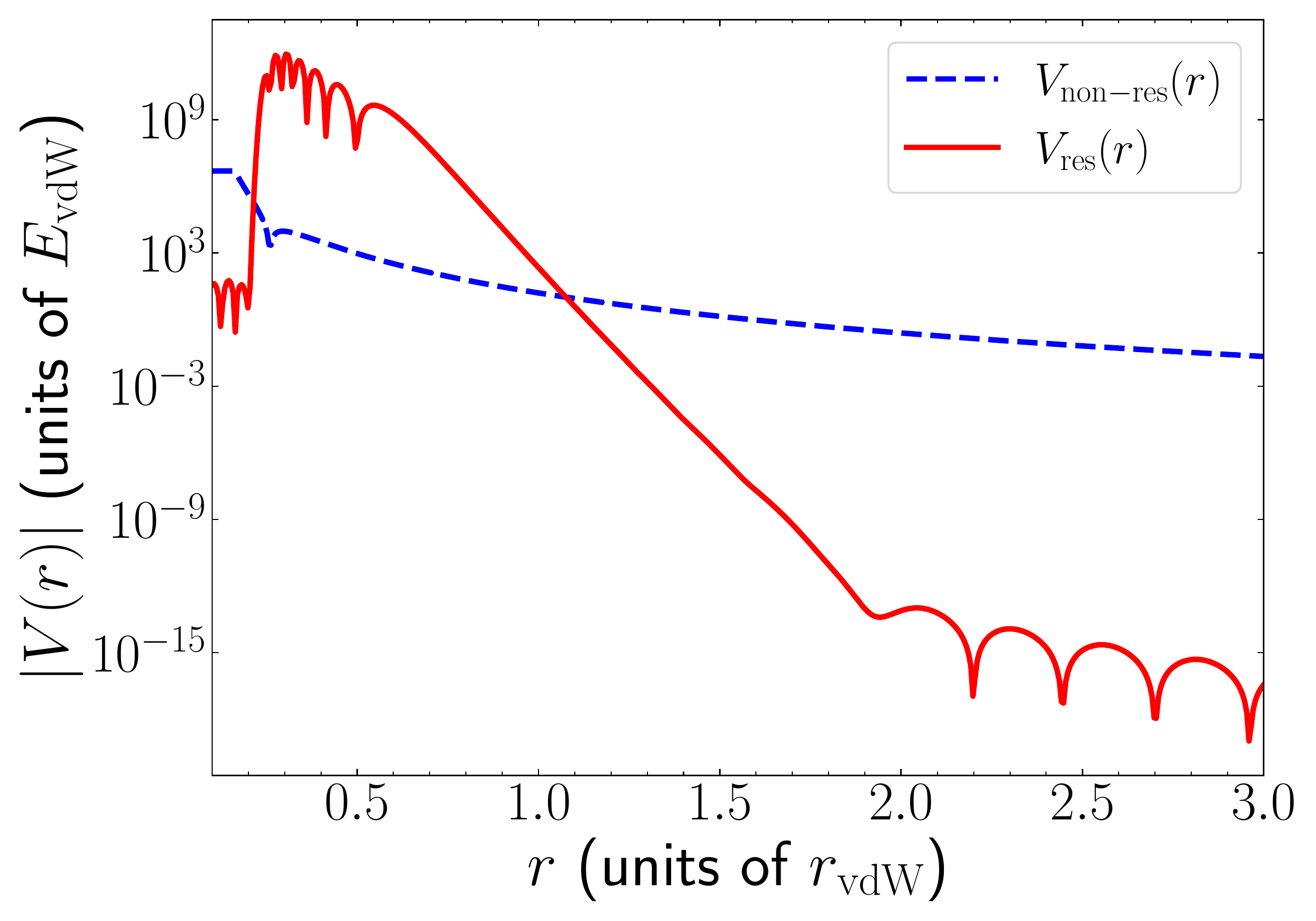}
\caption{Comparison of the absolute magnitude of the resonance enhanced interaction $V_{\mathrm{res}}(r)$, with a non-resonant van der Waals interaction $V_{\mathrm{non-res}}(r)$. Note that for $V_{\mathrm{res}}(r)$ the diagonal elements of the non-local potential matrix are plotted. Energy given in units of $E_{\mathrm{vdW}} = \hbar^2/(mr_{\mathrm{vdW}}^2)$}
\label{fig:Vcomp}
\end{figure}

The fixed spectating spin reduction of the multichannel STM equation, as discussed in Sec. \ref{sec:ESC_EST}, rests on the assumption that interactions in the non-resonant closed channels $\ket*{\underline{ab},c}$ and $\ket*{\underline{ac}, b}$ are negligible compared to the resonance enhanced channel $\ket*{\underline{bc}, a}$. To illustrate numerically that this is a reasonable assumption, we parametrise the effective interaction in the $\ket*{\underline{bc}}$ channel by computing an EST separable potential $V_{\mathrm{res}}$, associated with the resonant two-body bound state $\ket*{\varphi_{\mathrm{res}}}$. It is formulated as,
\begin{align}
\begin{split}
V_{\mathrm{res}} = V_{\underline{bc}, \underline{bc}} \ket*{\varphi_{\mathrm{res}}} \matrixel*{\varphi_{\mathrm{res}}}{V_{\underline{bc}, \underline{bc}}}{\varphi_{\mathrm{res}}}^{-1}\bra*{\varphi_{\mathrm{res}}}V_{\underline{bc}, \underline{bc}}
\end{split}
\end{align}
In Fig.~\ref{fig:Vcomp} we compare the magnitude of $\matrixel*{\vb{r}}{V_{\mathrm{res}}}{\vb{r}}$ to the van der Waals interaction without any resonant enhancement $V_{\mathrm{non-res}}$. One clearly observes that the resonant enhancement augments the bare interaction by several orders of magnitude, which supports the assumption made in Sec. \ref{sec:ESC_EST}.

\section{Comparison with nuclear interactions}
\label{ap:short}

\begin{figure}[t]
\centering
\includegraphics[height=0.19\textwidth]{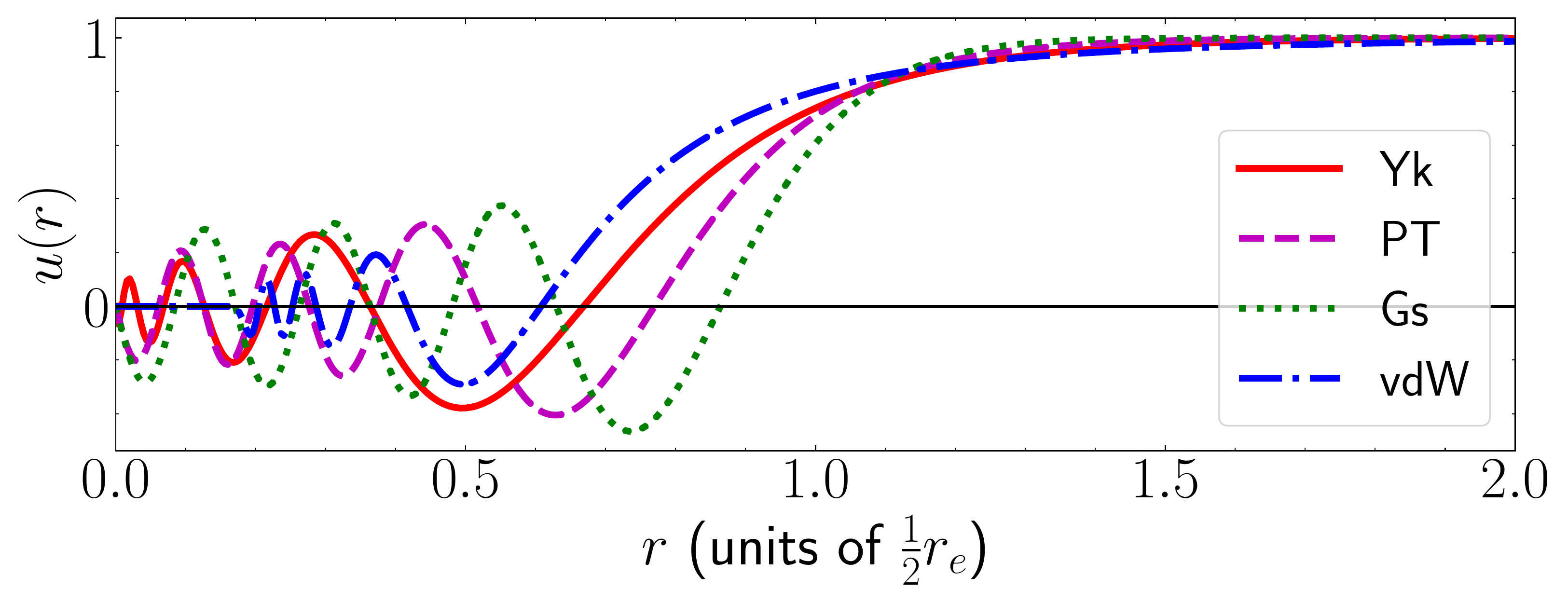}
\caption{Two-body radial wave functions at the appearance of the eighth potential resonance of the three nuclear interactions in Eq. \eqref{eq:modelpotshort} and the van der Waals (vdW) interaction in Eq. \eqref{eq:Vo}. All distances have been expressed in the associated effective range scale $r_e/2$.}
\label{fig:2body_short}
\end{figure}

In Sec. \ref{sec:SpinAn} we argue that our observations in the multichannel model arise due to a distinct interplay between the different possible spin-states on the three-body level. It is known however that the value of $\abs{a_-}$ can also increase with different kinds of asymptotic three-body interactions, without any need for additional spin-channels. In this appendix we contrast the changing three-body potential in systems of this type with our previous multichannel results, to show that the underlying mechanisms are indeed fundamentally different. Specifically we will consider single-channel EST models based on two-body interactions that decay exponentially in the long range, common in nuclear physics.  For this class of interactions the short-range two-body suppression looks rather different from the van der Waals potential, where "short-range" is now interpreted as $r < r_e/2$, with $r_e$ the effective range constant. To illustrate the differences we plot in Fig.~\ref{fig:2body_short} the two-body wave function at unitarity for the following set of nuclear potentials,
\begin{align}
\begin{split}
V_{\mathrm{PT}} &\sim - \eta\cosh^{-2}\left(r \right), \\
V_{\mathrm{Yk}} &\sim -  \frac{\eta}{r} \exp\left(-r\right), \\
 V_{\mathrm{Gs}}  &\sim - \eta \exp\left(-r^2 \right),
\end{split}
\label{eq:modelpotshort}
\end{align}
\begin{figure}[t]
\centering
\includegraphics[width = 0.45\textwidth]{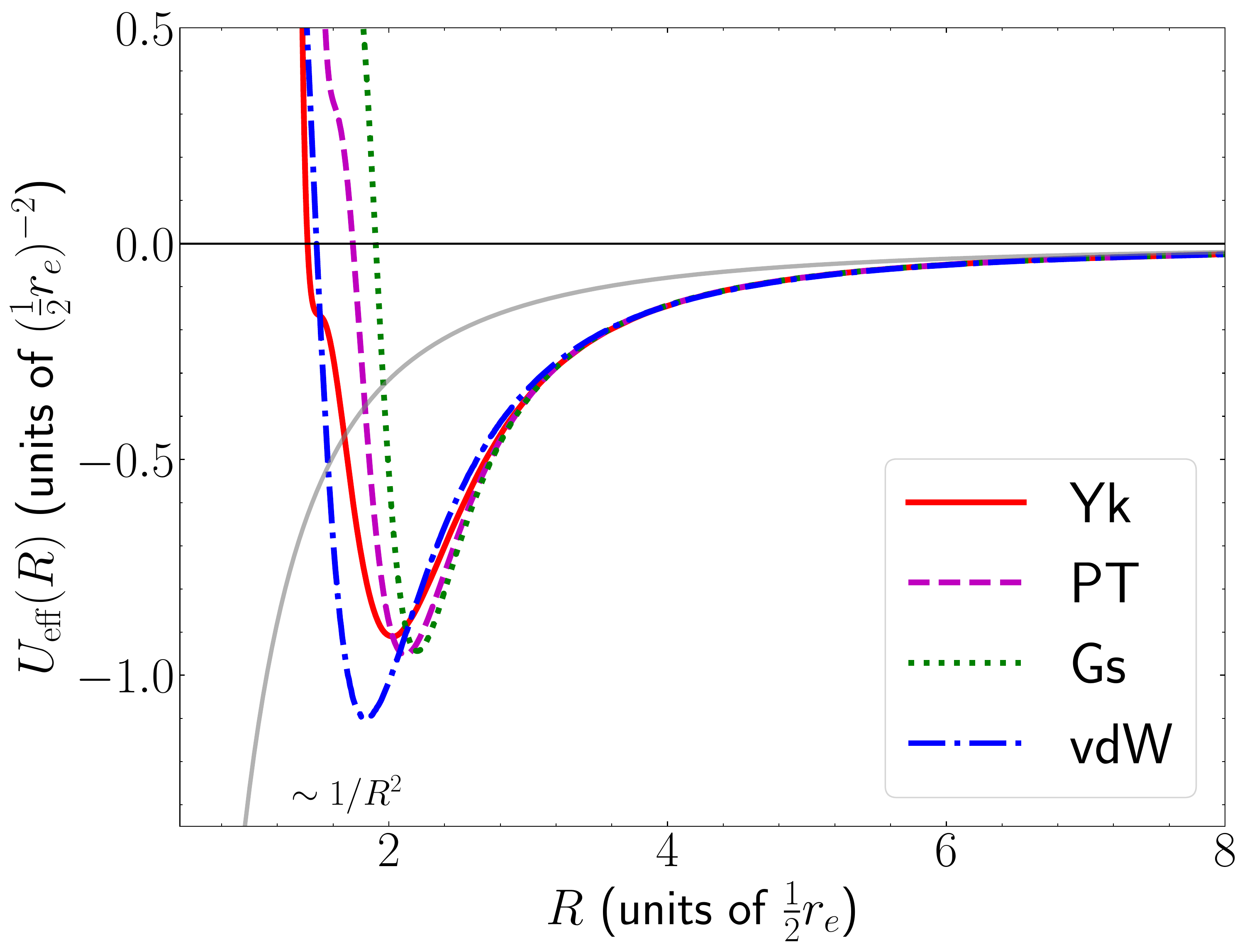}
\caption{Plot of the effective three-body potential as a function of the hyperradius, in units of the effective range. Colored lines show the effective potential for different shapes of the two-body interaction, tuned to a depth of eight dimer states. The grey line shows the asymptotic $\sim 1/R^2$ Efimov attraction as follows from the zero-range theory.}
\label{fig:EfPot_short}
\end{figure}

which respectively are the usual Pöschl-Teller (PT), Yukawa (Yk), and Gaussian (Gs) potentials \cite{Wang2012, Naidon2014_2}. The strength parameter $\eta$ is used to tune the potential towards resonance. Similar to the case of van der Waals interactions, the short-range suppression in the two-body wave function leads to the formation of a three-body repulsive barrier and hence a universal value for the three-body parameter in the limit of broad Feshbach resonances. As was shown in Ref. \cite{Naidon2014_2}, this universal value matches the three-body parameter that one obtains when using a simple step function as input into the EST model, which is zero for $r < r_e/2$ and unity everywhere else. Indeed for an infinite number of two-body bound states the PT, Yk and Gs wave functions in Fig.~\ref{fig:2body_short}, will all approach step functions \cite{Naidon2014_2}. This is not the case for the interaction in Eq. \eqref{eq:Vo}, whose infinitely deep limit is obtained by taking $r_0 \rightarrow 0$, yielding a pure (but ill-behaved) van der Waals potential. Interesting for this work is the fact that the three-body parameter $\abs*{a_-}$ obtained from the step-function limit in a single-channel model is $\sim 19 \ \%$ larger than the value obtained from the van der Waals interaction \cite{Naidon2014_2}. In our multichannel model we similarly observe an increased value of $\abs*{a_-}$ when $s_{\mathrm{res}}$ is decreased away from the broad resonance limit. Evidently however the associated change in the two-body wave function is very different, as becomes clear when comparing Figs.~\ref{fig:2body_sr} and \ref{fig:2body_short}. Whereas in the multichannel model the open-channel component that underlies the Efimov state was unchanged, for the interactions in this section there is a clear change in the short-range suppression in the open-channel. As shown in Fig.~\ref{fig:EfPot_short}, this subsequently leads to a shift in the location of the three-body barrier, which is absent in the multichannel model as explained in Sec. \ref{sec:SpinAn}.

\bibliography{References}

\begin{thebibliography}{56}
\bibitem [{LJP()}]{LJPexp}%
  \BibitemOpen
  \href@noop {} {}\bibinfo {note} {The advantage of using the $r^{-10}$ core
  instead of the more traditional $r^{-12}$ shape is the fact that the
  associated zero-energy scattering problem allows for an exact solution in
  terms of hyperconfluent geometric functions. Thus the length scale $r_0$ can
  be analytically related to the scattering length, making it straightforward
  to fix $a_{\mathrm{bg}}$. This interaction was characterised by J. Pade,
  cited in the next reference.}\BibitemShut {Stop}%
\end{thebibliography}

\end{document}